\documentstyle[aps,prb,epsf,twocolumn]{revtex}

\begin{document}

\twocolumn[\hsize\textwidth\columnwidth\hsize
\csname @twocolumnfalse\endcsname

\title{Critical Transverse Forces in Weakly Pinned Driven Vortex
Systems}

\author{Hans Fangohr$^{*\dag}$, Peter A. J. de Groot$^{\dag}$, Simon
J. Cox$^{*}$}

\address{$^{*}$Department of Electronics and Computer Science, \\
$^{\dag}$Department of Physics and Astronomy, \\ University of
Southampton, Southampton, SO17 1BJ, United Kingdom}

\maketitle

\begin{abstract}

We present simulation results of the moving Bragg glass r\'{e}gime of
a driven two-dimensional vortex system in the presence of a smoothly
varying weak pinning potential. We study the critical transverse force
and (i) demonstrate that it can be an order of magnitude larger than
previous estimates (ii) show that it is still observable when the
system is driven along low higher-order lattice vectors. We confirm
theoretical predictions that the critical transverse force is the
order parameter of the so-called moving glass phase, and provide data
to support experimentalists verifying the existence of a critical
transverse force.

\end{abstract}

\pacs{74.60Ge}

]

\narrowtext

\section{Introduction}
The vortex state is dominated by the competition of ordering and
disordering interactions. Vortex-vortex repulsion tends to order the
system whereas thermal fluctuations and pinning from material
imperfections introduce disorder into the vortex lattice. Recently,
interest has developed in the nature of the non-equilibrium states and
dynamical phases in the presence of a Lorentz force driving the
system. There is evidence from experiments,\cite{exp}
simulations \cite{Koshelev94,Jensen88,Moon96,Ryu96,Olson98} and 
theory \cite{Koshelev94,Giamarchi96,LeDoussal98,Balents98} that for
small driving forces the vortex system is disordered and shows
turbulent plastic flow, and that for larger driving forces the system
orders and shows elastic flow. For the ordered system Koshelev and
Vinokur \cite{Koshelev94} proposed that the vortices may form a moving
hexagonal crystal. Subsequently, Giamarchi and Le Doussal
\cite{Giamarchi96,LeDoussal98} predicted that this highly driven phase
may be a topologically ordered moving glass (the moving Bragg glass)
in which vortices move in elastically coupled static channels like
beads on a string. It was also suggested
\cite{LeDoussal98,Balents98,Scheidl98b} that the motion of vortices in
different channels may be decoupled (the moving transverse glass) and
thus shows smectic order. In computer simulations
\cite{Moon96,Ryu96,Olson98} and in experiments \cite{Pardo98andTroyanovski99}
both the
moving transverse glass (MTG) with decoupled channels and the moving
Bragg glass (MBG) with coupled channels have been observed.

A remarkable property of the moving glass (with either coupled or
de-coupled channels) is that, in the presence of random pinning and
once the static channels are established, the application of a small
force transverse to the direction of motion does not result in
transverse motion.\cite{Giamarchi96,LeDoussal98} Only if a critical
transverse force has been exceeded, is the system transversely
de-pinned. Computer simulations have demonstrated the existence of
such a critical transverse force for random
pinning,\cite{Moon96,Ryu96,Kolton99,Olson00b} and for periodic
pinning.\cite{Reichhardt99bandMarconi99}

In this work we use a more realistic representation of high purity
single crystals used in fundamental studies of vortex dynamics; we
investigate r\'{e}gimes with a high density of vortices with
long-range logarithmic vortex-vortex interaction potentials (as in
Ref. \onlinecite{Kolton99}) and we employ a weak smoothly varying
pinning potential rather than many strong point-like pins.
\cite{Moon96,Ryu96,Kolton99,Olson00b} We find and explain 
a magnitude of the critical transverse force of the order 
of 10\% of the static de-pinning force in the r\'{e}gime investigated 
in contrast to previous
works \cite{Moon96,Ryu96,Olson00b} which report it to be $\approx
1\%$. We report on novel results for the critical transverse force in
the presence of weak pinning which (i)~verify the theory of Giamarchi
and Le Doussal \cite{Giamarchi96} and (ii)~provide the first numerical
data which may be compared directly with current experimental efforts
to demonstrate the existence of the critical transverse force.

\section{The simulation}
\begin{figure}
\centerline{\epsfxsize=6cm \epsfbox{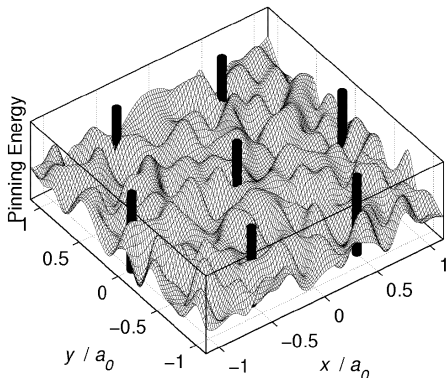} }
\caption{A sample pinning potential. Distances in $x$- and
$y$-directions are given in multiples of the vortex lattice spacing,
$a_0$. The seven black cylinders indicate vortex lines separated by
$a_0$ to demonstrate the length scale.
\label{figure1}}
\end{figure}

We model the vortex motion of a two-dimensional system with over-damped
Langevin dynamics. The total force acting on vortex $i$ is given by
${\bf F}_i= -\eta {\bf v}_i + {\bf F}^{\text{L}} + {\bf
  F}_i^{\text{vv}} + {\bf F}_i^{\text{vp}} + {\bf F}_i^{\text{therm}}
= \bf{0}$ where $\eta$ is the viscosity coefficient, ${\bf v}_i$ the
velocity, $ {\bf F}^{\text{L}}$ the Lorentz force, ${\bf
  F}_i^{\text{vv}}$ the vortex-vortex interaction, ${\bf
  F}_i^{\text{vp}}$ the vortex-pinning interaction, and ${\bf
  F}_i^{\text{therm}}$ a stochastic noise term to model temperature.
The vortex-vortex interaction force appropriate for rigid vortices in
thin films and pancakes in decoupled layers of layered materials
is\cite{Clem91} $ {\bf F}_i^{\text{vv}}= (\Phi_0^2 s)
(2\pi\mu_0\lambda^2)^{-1} \sum_{j \ne i} ({\bf r}_i-{\bf r}_j) (|{\bf
  r}_i-{\bf r}_j|)^{-2}$.  $\Phi_0$ is the magnetic flux quantum,
$\mu_0$ the vacuum permeability and $s$ the length of the vortex. We
employ periodic boundary conditions and cut off the logarithmic
vortex-vortex repulsion potential at half the system size. It is
important to reduce the vortex-vortex interaction near the cut-off
distance smoothly to zero.\cite{Fangohr00c} We investigate systems
with a magnetic induction of $B=1 \text{ T}$ and a penetration depth
of $\lambda = 1400 \text{\AA}$ which yields a vortex density of
$\approx 10/\lambda^2$. The random pinning potential as shown in Fig.\ 
\ref{figure1} varies smoothly on a length scale of $\lambda/25$ which
is of the order of magnitude of the coherence length $\xi$. The root
mean square value of the corresponding pinning forces is denoted by
$F^{\text{vp}}_{\text{rms}}$.  System sizes from 100 to 3000 vortices
have been investigated. Forces are expressed in units of the force,
$f_0$, that two vortices separated by $\lambda$ experience.

\begin{figure}
\centerline{\epsfxsize=6cm \epsfbox{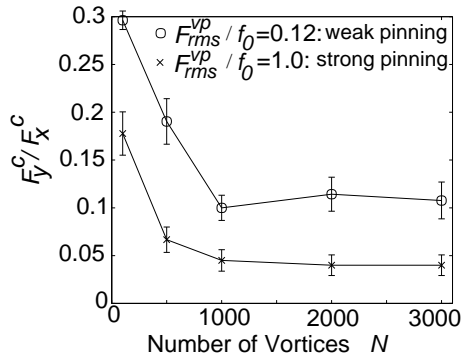}}\bigskip
\caption{The ratio of the critical transverse force $F^c_y$ to the
  static de-pinning force $F^c_x$ for various numbers of vortices,
  $N$. The lower curve is for strong pinning with
  $F^{\text{vp}}_{\text{rms}}/f_0=1.0$ and the upper curve is for weak
  pinning with $F^{\text{vp}}_{\text{rms}}/f_0=0.12$.
\label{figure2}}
\end{figure}

Initially, we anneal the vortex system in the presence of random
pinning from a molten state to zero temperature.  Then a driving force
is applied which is increased every $4\cdot10^{4}$ time steps. With
increasing driving force we find a pinned system, turbulent plastic
flow and finally the MBG. For sufficiently strong pinning there is an
intermediate r\'{e}gime between turbulent plastic flow and the MBG in
which the vortex motion in different channels is decoupled.
\cite{Fangohr00b} We find a critical transverse force for both the MBG
and the MTG, and here we report on the small pinning strengths which
do not allow smectic states with decoupled motion of channels of
vortices. To find the critical transverse force we start with a MBG
driven by a constant force $F^{\text{L}}_x$ in the $x$-direction and
slowly increase the transverse force $F_y^{\text{L}}$ in the
$y$-direction, until the system starts moving transversely. The lower
ends of the bars shown in Fig.\ \ref{figure2} to \ref{figure5}
represent the largest probed transverse force which did not yield any
transverse motion, and the upper ends of the bars show the smallest
transverse force that could de-pin the system transversally.

\section{Results}

Fig.\ \ref{figure2} shows that there is a decrease in the ratio of the
critical transverse force $F^c_y$ to the static de-pinning force
$F^c_x$ for system sizes below 1000 vortices. However, for larger
systems this ratio remains constant, showing that the observed $F^c_y$
is not a finite-size effect. We have increased the cut-off with the
system size to ensure that effects due to the long-range interactions
between the additional particles in the simulation are taken into
account, which contrasts to a similar finite-size study
\cite{Olson00b} where the cut-off for the vortex-vortex interaction
was kept constant and the results were reported to be independent of
the system size.
 
Previous estimates \cite{Moon96,Ryu96,Olson00b} for the ratio
$F^c_y/F^c_x$ give a value $\approx 0.01$. We find $F^c_y/F^c_x
\approx 0.1$ and identify two reasons for this order of magnitude
discrepancy.  Firstly, we study the weak pinning r\'{e}gime in which
the hexagonal structure of the \emph{static} vortex system (\emph{i.e.}
without an applied driving force) is not completely destroyed due to
the vortex pinning, whereas previous studies focused on the strongly
pinned r\'{e}gime in which the \emph{static} system is strongly
disordered. Both systems --- with weak and strong pinning --- move
elastically and show topological order under the influence of the
driving force in the $x-$direction. The critical force required to
de-pin the static system, $F^c_x$, is greater for the strongly pinned
system which shows disorder, because a disordered system can adopt
better to the pinning potential. However, the force required to de-pin
the moving system transversely, the critical transverse force $F^c_y$,
depends less strongly on the pinning strength because the elastically
moving system is topologically ordered for either pinning strength.
Thus, the ratio $F^c_y/F^c_x$ is higher for weak pinning. We
demonstrate this in Fig.\ \ref{figure2} where we show that the change
from strong to weak pinning increases the ratio $F^c_y/F^c_x$ by a
factor 2 to 3.  Secondly, the pinning potentials employed in Refs.
\onlinecite{Moon96,Ryu96} and \onlinecite{Olson00b} consist of
(strong) point-like randomly distributed pins, which we find increase
the static de-pinning force $F^c_x$ by another factor 2 to 3 compared
with using a smoothly varying pinning potential (Fig.\ \ref{figure1}).
We would thus get to the same order of magnitude for the ratio
$F^c_y/F^c_x$ as Refs. \onlinecite{Moon96,Ryu96} and
\onlinecite{Olson00b} if we used the simulation scenario they
employed. We find that for different random pinning configurations the
$F^c_y$ can vary up to a factor 2 in the weak pinning limit.

\begin{figure}
\centerline{\epsfxsize=8cm \epsfbox{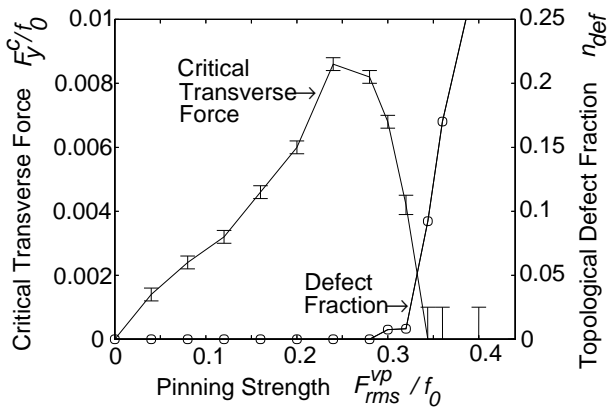}}\bigskip
\caption{Critical transverse force and topological defect fraction
$n_{\text{def}}$ as a function of pinning strength. The critical
transverse force reduces to zero where the system changes from elastic
flow to turbulent plastic flow and the number of topological defects
increases rapidly.
 \label{figure3}}
\end{figure}

Fig.\ \ref{figure3} shows the variation of $F_y^c$ as a function of
the pinning strength for systems driven with a constant driving force
$ F_x^{\text{L}}=0.3 f_0$ in the $x$-direction. The absence of
transverse barriers for zero pinning strength shows that it is not the
periodic boundary conditions which result in a critical transverse
force. With increasing pinning strength $F_y^c$ increases linearly
until it starts to decay for pinning strengths of
$F^{\text{vp}}_{\text{rms}}\approx 0.25 f_0$ and reaches zero at
$F^{\text{vp}}_{\text{rms}}\approx 0.35 f_0$. The decay of the $F_y^c$
is caused by the strength of the pinning producing turbulent plastic
flow of the vortices: in this region the MBG breaks down. This is
demonstrated by the second curve in Fig.\ \ref{figure3} which shows
that the fraction of vortices that are topological defects,
$n_{\text{def}}$, increases rapidly for pinning strengths greater than
$0.325 f_0$. We define a topological defect to be one which does not
have six nearest neighbors in the periodic Delaunay triangulation of
the vortex positions. The slight increase of $n_{\text{def}}$ for
pinning strengths $0.3 f_0$ and $0.325 f_0$ is due to strong temporary
deformations of the MBG such that pairs of topological defects appear
next to each other and disappear after a few time steps. This
indicates the weakness of the MBG but not its breakdown (because the
system shows elastic motion). In contrast, the transition to turbulent
plastic flow is accompanied by a proliferation of topological defects.
This confirms theoretical expectations \cite{LeDoussal98} that the
critical transverse force, $F_y^c$, is the order parameter for the
moving glass, which, in the weak pinning r\'{e}gime, is represented by
the MBG. The data shown in Fig.\ \ref{figure3} are obtained for a
system of 576 vortices. For larger systems we get qualitatively the
same curves, with a slightly reduced height of $F_y^c$.

\begin{figure}
\centerline{\epsfxsize=6cm \epsfbox{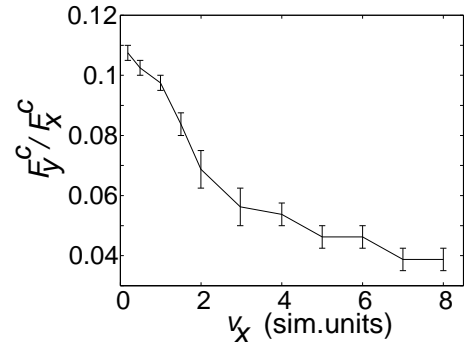}} \bigskip
\caption{Critical transverse force $F_y^c$ normalized by the static
de-pinning force $F_x^c$ for various longitudinal velocities $v_x$ in
the moving Bragg glass r\'{e}gime.
\label{figure4}}
\end{figure}

Le Doussal and Giamarchi\cite{LeDoussal98} suggested a dependence of
the critical transverse $F^c_y$ as a function of the longitudinal
velocity, $v_x$, which predicts a decay of the $F_y^c$ for large $v_x$
and has not previously been investigated numerically.  For an
isotropic system one expects that the critical ``transverse'' force
$F^c_y$ for a static system is the same as the critical force (acting
in any direction) that is required to de-pin the system. Our computations
confirm that in particular $F^c_y=F^c_x$ for a static system. However,
as soon as the system of vortices is de-pinned and moves elastically
in the $x$-direction, the transverse critical force reduces to much
smaller values because the system is not sticking to the pinning
potential, but de-pinned in the $x$-direction. 
 Fig.\ \ref{figure4} shows results of our simulations using a pinning
strength of $F^{\text{vp}}_{\text{rms}}=0.12 f_0$ and a system size of
1200 vortices. We could not resolve the smallest velocities because
these are computationally expensive, and we have omitted the data
point at $v_x=0$. The curve starts for small $v_x$ with a $F^c_y$ of
$\approx 10\%$ of the static de-pinning force $F^c_x$. With increasing
$v_x$ the $F_y^c$ decreases quickly up to velocities of $\approx 2$
simulation units and then less strongly for larger velocities. Our
findings are compatible with the prediction that the critical
transverse force decays for higher velocities as additional dynamic
disorder weakens the transverse barriers.\cite{LeDoussal98}

The velocities given here in simulation units are directly comparable
to Ref. \onlinecite{Fangohr00b}, where 0.1 represents a low velocity
for the system, and 10 is large. In these units the transition from
plastic to elastic flow happens around a velocity of 7 simulation
units with a pinning strength of $F^{\text{vp}}_{\text{rms}}= 1.0
f_0$. Our work suggests therefore that small driving forces are most
appropriate for experimental verification of the critical transverse
force.

It is worth noting that using a system size of less than 1000 vortices
(Fig.\ \ref{figure2}) gives qualitatively different results; for such
small systems the $F_y^c$ in Fig. \ref{figure4} remains constant above
a small velocity of $\approx$ 0.3 simulation units, which is a
finite size effect.

\begin{figure}
\centerline{\epsfxsize=7cm \epsfbox{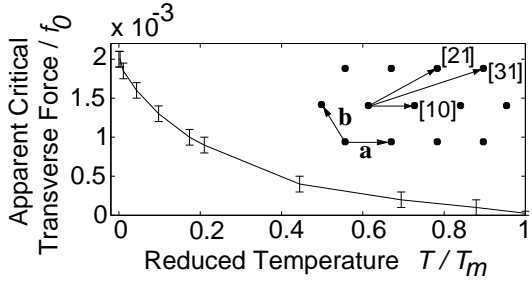}}\bigskip
\caption{Apparent critical transverse force as a function of reduced
temperature $T/T_m$. Inset: Directions of the driving force probed for
the existence of static channels and critical transverse forces. The
lattice vectors used for labeling the directions are shown as ${\bf
a}$ and ${\bf b}$. \label{figure5}}
\end{figure}

\begin{figure}
\centerline{\epsfxsize=8cm \epsfbox{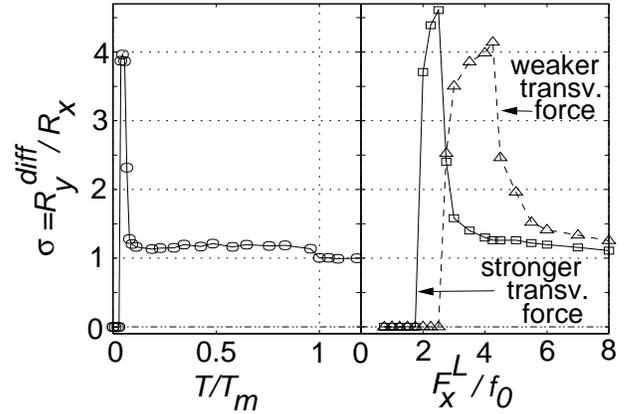}}\bigskip
\caption{Differential resistance in transverse direction normalized by
resistance in longitudinal direction as a function of reduced temperature
({left}) and of strength of the longitudinal driving force
({right}). See text for details.
\label{figure6}}
\end{figure}

We report on the existence of the critical transverse force in higher
commensuration directions. The data shown in Fig.\ \ref{figure2} to
\ref{figure6} are obtained with a driving force acting in the [10] (or
equivalent symmetry) directions of the Bragg-Glass-lattice (see inset
Fig.\ \ref{figure5}). The theory of Giamarchi and Le~Doussal
\cite{Giamarchi96} predicts that the system should see static disorder
when moving in any commensurate direction. It is then expected that
the channels and transverse pinning should exist for low
commensuration vectors and become unstable at higher ones due to the
relatively increasing dynamic disorder.\cite{Giamarchi96} To test
these ideas, we have applied a driving force to a hexagonal lattice in
the [21]- and [31]-direction. For the [21]-direction we observe that
static channels characteristic of the MBG establish and that there are
transverse barriers to a transverse force which is subsequently
applied. In contrast, we have found that for the [31]-directions
static Bragg channels do not develop. We presume them to be unstable
(at these velocities), and consequently, no critical transverse force
has been found for the [31]-direction.

For finite temperatures it is predicted that there is no true critical
transverse force but all transverse drives result in a small response
in the transverse motion of the system.\cite{LeDoussal98} However, for
an apparent critical transverse force the system is expected to start
moving transversely much quicker. Data on the apparent critical
transverse force in Fig.\ \ref{figure5} shows that it decays with
increasing temperature and vanishes at the melting temperature of the
system. The de-pinning force of the static system, $F^c_x$, decays
similarly with increasing temperature, such that $F^c_x/F^c_y \approx
\mathrm{const}$.

To assist in the experimental demonstration of the existence of the
critical transverse force we provide in Fig.\ \ref{figure6} data on
the differential transverse resistance $R^{\text{diff}}_y={\mathrm{d}}
v_y/{\mathrm{d}} F^{\text{L}}_y$ normalized by the longitudinal
resistance $R_x={v_x}/{F^{\text{L}}_x}$, which can both be measured
experimentally.  We use central differences to approximate the
differential transverse resistance $$R^{\text{diff}}_y =
\frac{{\mathrm{d}} v_y}{{\mathrm{d}} F^{\text{L}}_y} \approx
\frac{v_y(F^{\text{L}}_y+\Delta)-v_y(F^{\text{L}}_y-\Delta)}{2\Delta}
$$
where $\Delta$ is a small change in force. We compute $\sigma =
R^{\text{diff}}_y/R_x$, which is a function of temperature, $T$, and
both components, $F^{\text{L}}_x$ and $F^{\text{L}}_y$, of the driving
force: $\sigma=\sigma(T, F^{\text{L}}_x, F^{\text{L}}_y)$. We choose a
small transverse force, $F^{\text{L}}_y$, and keep it constant for
each curve in Fig.\ \ref{figure6}. In the left plot we show $\sigma(T,
F^{\text{L}}_x\!\!=\!\!1f_0, F^{\text{L}}_y\!\!=\!\!0.00225f_0)$,
\emph{i.e.} we vary the temperature $T$. And in the right plot we show
two curves with slightly different transverse forces at zero
temperature: $\sigma(T\!\!=\!\!0, F^{\text{L}}_x,
F^{\text{L}}_y\!\!=\!\!0.00225f_0)$ and $\sigma(T\!\!=\!\!0,
F^{\text{L}}_x, F^{\text{L}}_y\!\!=\!\!0.00275f_0)$, \emph{i.e.} we
vary the longitudinal driving force $F^{\text{L}}_x$.
 
In the left plot in Fig.\ \ref{figure6} the constant transverse force
$F^{\text{L}}_x=1f_0$ results in a velocity of $v_x \approx 1$
simulation units, and the transverse force $F^{\text{L}}_y\!=\!0.00225
f_0$ is chosen to be slightly smaller than the critical transverse
force at $T=0$ for these simulations. The plot shows that for very
small temperatures $\sigma \approx 0$. This means that an increase in
the transverse force does not result in an increase in transverse
motion. With increasing temperature $\sigma$ shows a peak. Here, an
increase in the transverse force results in a strong increase in the
transverse velocity, and this is where the system starts quickly
moving transversely. For a further increase in temperature, $\sigma$
comes down to $\sigma \approx 1.2$, before it drops to 1.0 at the
melting temperature $T_m$. The reason that $\sigma \approx 1.2$ for
intermediate temperatures is that even after transverse de-pinning the
moving system feels some transverse pinning up to transverse forces
many times larger than the critical transverse force.\cite{Olson00b}
The remaining transverse pinning reduces with increasing transverse
drive, $F_y^{\text{L}}$, and we find the transverse response, $v_y$,
to be non-linear in this regime: $v_y$ increases stronger than
linearly with $F_y^{\text{L}}$. Thus, $\sigma =
(\text{d}v_y/\text{d}{F_y^{\text{L}}})/R_x > 1$.

The right plot in Fig.\ \ref{figure6} shows zero temperature data for
various longitudinal driving forces $F^{\text{L}}_x$ and two different
constant transverse forces $F^{\text{L}}_y\!=\!0.00225 f_0$ and
$F^{\text{L}}_y\!=\!0.00275 f_0$. For small $F^{\text{L}}_x$ the
system does not move transversely and $\sigma=0$.  When
$F^{\text{L}}_x$ increases, it increases the velocity $v_x$ of the
system and thus reduces the critical transverse force (as shown in
Fig.\ \ref{figure4}). Therefore, for sufficiently large
$F^{\text{L}}_x$ the system starts moving transversely and $\sigma$
shows a peak which decays to 1.0 for larger $F^{\text{L}}_x$. The slow
decay of $\sigma$ is due to remaining transverse pinning above the
transverse de-pinning force.\cite{Olson00b} The magnitude of the
constant transverse driving force $F^L_y$ determines the position of
the peak of $\sigma$, as the two curves in the right plot in Fig.\ 
\ref{figure6} demonstrate. In experimental work the presence of a
critical transverse force should manifest itself in $\sigma$ changing
as shown in Fig.\ \ref{figure6}.

\section{Summary}  

We have investigated numerically the critical transverse force of
two-dimensional vortex systems in the presence of a random pinning
potential. We find a critical transverse force for both the MBG and
the MTG, but not for turbulent plastic flow. The ratio of the critical
transverse force to the static de-pinning force is of the order of
10$\%$.  For the MBG we find that the critical transverse force
increases with increasing pinning strength up to a value at which the
elastic motion changes to turbulent plastic flow and the critical
transverse force goes rapidly to zero. The critical transverse force
is inversely proportional to the longitudinal velocity and is
compatible with theoretical predictions.\cite{LeDoussal98}

We have performed simulations in which a hexagonal lattice is driven
in low higher-order lattice directions. These simulations revealed for
the first time that a MBG and a critical transverse force exist for
the driving force in the [21]-direction, but not for the [31]- and
higher-order directions, thus supporting the theory of Giamarchi and
Le Doussal.\cite{Giamarchi96} Our results suggest that in an
experimental search for the critical transverse current low
temperatures and small longitudinal driving forces (which however will
have to be large enough to cause elastic motion) are most
promising. We provide data that can be compared directly with
experimental efforts to demonstrate a critical transverse force.

\section{Acknowledgments} We thank P. Le Doussal, A. R. Price 
and S. Gordeev for helpful discussions. We acknowledge financial 
support from DAAD and EPSRC.

\end{document}